# THE PULSAR SEARCH COLLABORATORY

R. Rosen
National Radio Astronomy Observatory
P.O. Box 2, Green Bank, WV 24944
rrosen@nrao.edu

S. Heatherly, National Radio Astronomy Observatory
M. A. McLaughlin, West Virginia University
R. Lynch, University of Virginia
V. I. Kondratiev, West Virginia University
J. R. Boyles, West Virginia University
M. Wilson, West Virginia University
D. R. Lorimer, West Virginia University
S. Ransom, National Radio Astronomy Observatory

**Abstract**

The Pulsar Search Collaboratory [PSC, NSF #0737641] is a joint project between the National Radio Astronomy Observatory (NRAO) and West Virginia University (WVU) designed to interest high school students in science, technology, engineering, and mathematics [STEM] related career paths by helping them to conduct authentic scientific research. The 3-year PSC program, which began in summer 2008, teaches students to analyze astronomical radio data acquired with the 100-m Robert C. Byrd Green Bank Telescope for the purpose of discovering new pulsars. We present the results of the first complete year of the PSC, which includes two astronomical discoveries.

## 1. FOREWORD

High school freshman Lucas Bolyard was bored one Saturday. He decided to log on to the Pulsar Search Collaboratory [PSC] database and analyze some plots. He had become adept at the work after having scored over 2,000 of them in the few months since he joined his school's research team. Among the dozens of data sets he looked at that day, one single pulse plot intrigued him. Single pulse plots (one of two types of diagnostic plots used to analyze data in the database) display the properties of short, transient signals. Usually these signals are produced by radio frequency interference.

But what caught Lucas' s eye was a bright signal at a non-zero dispersion measure. This is a clue that the signal came from space, not Earth. So he reported it, and after a great deal of persistence on his part, it went on a list of candidates for astronomers to observe with the Green Bank Telescope [GBT]. Disappointingly, the follow-up observations showed nothing, indicating that the object was not a normal pulsar. And still Lucas asked: “What could cause such a signal?” Eventually astronomers went back to the original raw data and confirmed that Lucas had discovered a signal that was of astronomical origin and most likely a cosmic object called a rotating radio transient (McLaughlin 2006). Lucas’s discovery was announced in September 2009 precipitating an invitation to the White House and a meeting with the President and First Lady, during a Star Party event held on the White House lawn!

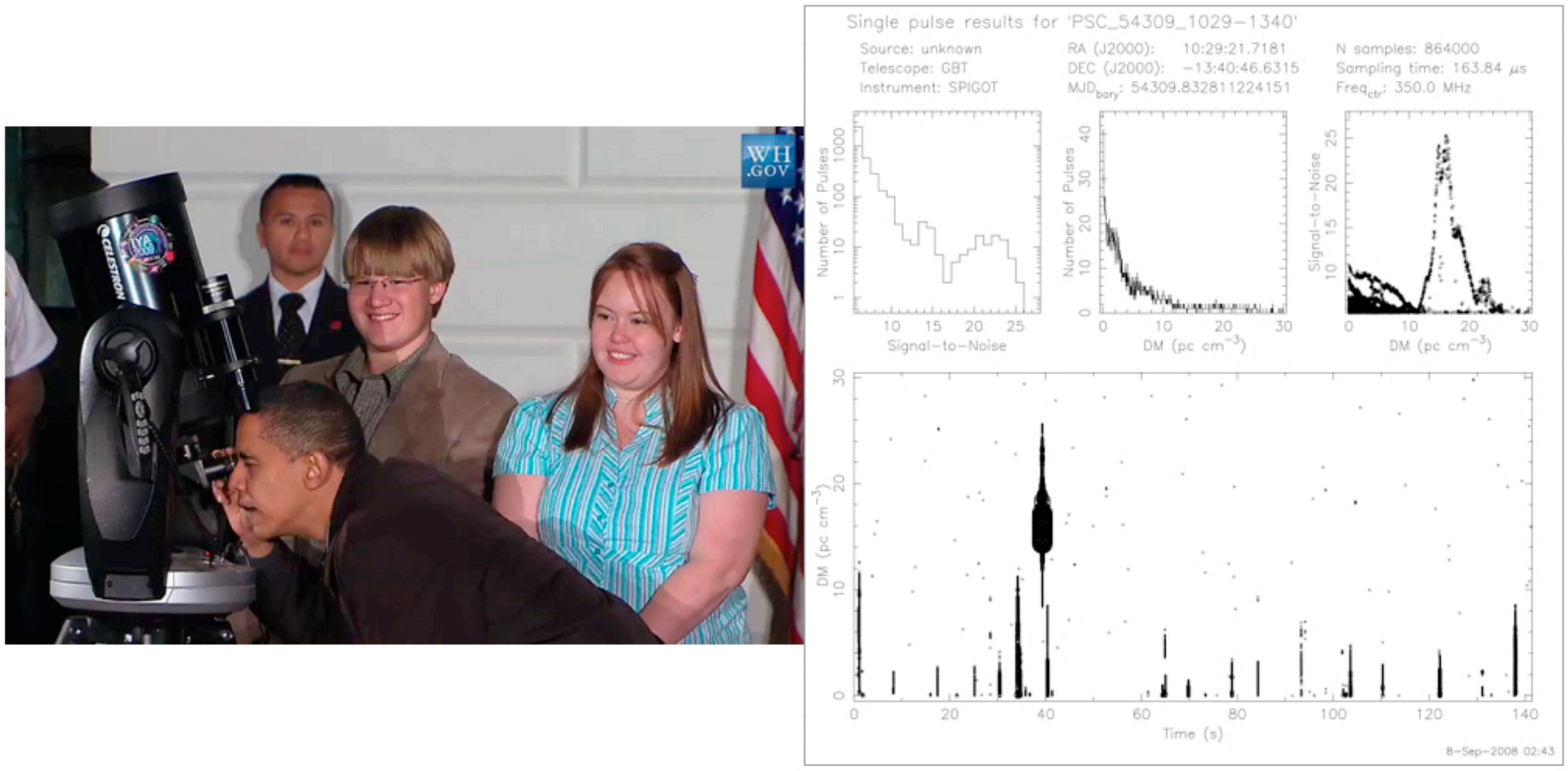

*Figure 1. Left: Lucas Bolyard of South Harrison High School, West Virginia, looks on as President Obama begins a star party at the White House. Right: The diagnostic single pulse plot in which Lucas found a radio burst.*

Lucas is a member of the Pulsar Search Collaboratory (http://www.pulsarsearchcollaboratory.com), a growing group of high-school students who have access to over thirty terabytes of data collected using the 100-m Robert C. Byrd Green Bank Telescope for the purpose of discovering new pulsars.

### 1.1 Overview of the Pulsar Search Collaboratory Program

The science, technology, engineering, and mathematics [STEM] related fields are crucial to America’s success in the growing global economy. While the demand for skilled technical people in the workforce is growing – 80% of the fastest growing fields in the United States are in science and technology areas (Coble & Allen 2005; Dorsen 2006) – the current workforce in these areas is shrinking due to retirement and lack of interest at the collegiate level (Dorsen *et al*. 2006). This may be in part because high-school teachers are not fully preparing students for STEM career pathways. High - school teachers, though they may agree with the theoretical tenets of the nature of science, do not adequately understand the nature of science (National Committee on Science Education Standards and Assessment & National Research Council 1996; Hollweg *et al*. 2003). Furthermore, while state curriculum standards may address the skills needed to prepare students for STEM career pathways, large gaps exist between those standards and the actual curriculum implemented by the teachers (Lederman 1992; Lynch 1997; Bybee 2000).

The National Science Foundation created the Innovative Technology Experiences for Students and Teachers [ITEST] program to address the shortages in STEM related fields by funding comprehensive projects for teachers and students in formal settings and out-of-school programs for students. The Pulsar Search Collaboratory is a comprehensive program, designed to build students' awareness of, and interest in, STEM careers.

As part of the extensive outreach program of the National Radio Astronomy Observatory [NRAO], the PSC is designed to interest high-school students in STEM related careers using radio astronomy data in a student research program that aims to find new pulsars – rapidly rotating highly magnetized neutron stars – in our Milky Way Galaxy.

The array of programs designed to involve the public and students in astronomical research is growing. The venerable SETI@home uses personal computers that are connected to the internet to analyze radio data, searching for extraterrestrial intelligence (http://setiathome.org). This program allows the public to participate by providing CPU cycles, but it is not interactive: There is no communication between the scientists and the participants.

Galaxy Zoo (http://www.galaxyzoo.org) places optical galaxy images from the Sloan Digital Sky Survey online where participants classify them. Over 150,000 people have participated in Galaxy Zoo, and the project has resulted in scientific publications (Darg 2010a; Darg 2010b). The interface to Galaxy Zoo's online database provides the necessary training for participants by helping them classify galaxies based upon the shapes seen in the image.

Other interactive programs include Stardust@Home and the International Asteroid Search Campaign. Like Galaxy Zoo, Stardust@Home (http://stardustathome.ssl.berkeley.edu) trains participants online to identify images of interstellar dust particles. The International Asteroid Search Campaign [IASC] has students search optical CCD images for asteroids, near earth objects, and supernovae (Miller 2008). The IASC (http://iasc.hsutx.edu) provides teacher training workshops and shows teachers how they can integrate the IASC into the classroom. Students learn about asteroids, transient phenomena (like supernovae), and the basics of optical astronomy, and have discovered five Main Belt asteroids.

What sets the PSC apart from these interactive programs is that students are working with radio data, not optical data. Furthermore, the PSC data are not images. Our diagnostic pulsar plots contain much more information than optical images. To extract science from the data, the students must understand the basics of observational radio astronomy, the nature of pulsars and their signature in the plots, and the sources of radio frequency interference that easily can bbe mistaken for a pulsar. A simple online classification scheme is insufficient. And unlike SETI@home, this level of understanding requires constant communication between the astronomers and students.

Two other programs, the Arecibo Remote Command Center [ARCC] program at the University of Texas, Brownsville, and the Pulse@Parkes Program, run by the Australia Telescope National Facility, also involve students in pulsar data analysis. ARCC students in Brownsville, Texas, search for new pulsars in data taken with the Arecibo telescope. This program (http://arcc.phys.utb.edu/ARCC) focuses on a local group of high-school students, and University of Texas Brownsville undergraduate students, and does not provide teacher instruction. Pulse@Parkes (http://outreach.atnf.csiro.au/education/pulseatparkes) is a spin-off of the AARC program and involves students from Australia and Europe in a program of monitoring known radio pulsars (Hobbs 2009).

The data collected for the PSC belong solely to the students. While the PSC is part of a larger collaboration of professional astronomers, these data have not been "pre-examined" by astronomers. Once raw data from the Green Bank Telescope are processed into diagnostic plots, the plots are completely analyzed by PSC students and the discoveries belong to them. The ownership of the data, the complexity of the plots for pulsar searching, and the interactive nature of the science make the PSC unique among student research programs.

Many of our colleagues, including reviewers of our grant proposals, express doubt about the viability of conducting radio astronomy outreach projects that involve students in authentic research. Reasons cited are that it is too hard, too esoteric, and not connected to school curricula. The PSC project is experimental and seeks to learn if these concerns can be addressed. After all, radio astronomy is responsible for some of the greatest astronomical discoveries in the Twentieth Century, including pulsars.

## 2. THE PULSAR SEARCH COLLABORATORY PROGRAM

The PSC began in July 2008, and we are currently in the middle of our second year. So far, 25 teachers from 24 schools have participated in our program. During the first year, 153 students joined the PSC, and we have gained 79 new students thus far in our second year; 232 students have participated in the PSC to date. Of the first year students, 43 are still active members who are currently inspecting plots. Our 122 currently active members (55 boys, 67 girls) come from seventeen schools in six states.

While the primary goal of the PSC is to stimulate student interest in STEM careers, other goals include: advancing high-school science teachers' and students' understanding of the nature of science and the relationship between science and technology, preparing teachers to implement authentic research with students, and promoting student use of information technologies.

To reach as many students as we can, while also providing in-person training, the PSC has adopted a novel organizational scheme: We train teachers and student leaders, who are chosen by the teachers, in radio astronomy and the techniques of pulsar searching, and then the teachers and student leaders take this information to their schools and recruit other interested students. Each year the PSC starts with a three-week workshop at NRAO in Green Bank, West Virginia, for teachers and student leaders. The teachers learn the fundamentals of astronomy, focusing on radio astronomy and pulsars, and devise whole-class activities to bring to the classroom. The student leaders, along with the teachers, learn the details of pulsar searching, observational radio astronomy, and radio frequency interference [RFI]. During the academic year, teachers recruit new students to participate in the PSC through the whole-class activities, then student leaders and teachers form teams at their schools to participate in the PSC throughout the year. The PSC culminates with a Capstone Seminar at West Virginia University [WVU], where the students learn about STEM fields on campus and present their research results.

## 2.1 Components of the Pulsar Search Collaboratory Program

### 2.1.1 *Summer Workshop in Green Bank*

#### *Teacher Institute*

To involve the teachers in research and provide them with the necessary foundation in astronomy, the teachers spend two weeks in Green Back at NRAO over the summer. The teachers learn about the fundamentals of astronomy through a mini-course, and research talks given by astronomers, and investigate simple research questions using a working 40-foot diameter radio telescope.

The entire Teacher Institute, and particularly the research experience, is a model for inquiry-based teaching and learning where the participants, in this case the teachers, assume the role of students. They conduct research projects using pulsar data, much as their students will, and they present their results to each other and PSC staff. Teachers design inquiry-based classroom activities that introduce students to radio astronomy and pulsar research while addressing core content standards. We devote the last two days of the Teacher Institute to intense planning for the Student Institute.

Teachers who complete the Teacher Institute are awarded three hours of graduate credit in Physics from WVU. Teachers receive an additional 3 hours of graduate credit in Curriculum and Instruction after completing academic year activities.

### *Student Institute*

Student leaders join their teachers at the NRAO for a six-day student orientation to the PSC. The Student Institute is an abbreviated version of the Teacher Institute. The week focuses on specialized topics: pulsars and RFI, observational techniques in radio astronomy, interpreting pulsar plots, and the use of the PSC website and database. Teachers practice their new knowledge by leading most of the activities and lessons.

As with the teachers the previous week, we present the students with a research problem: the task of searching for new pulsars. As part of their research experience, the students find and identify known pulsars and receive observing time on the Green Bank Telescope [GBT] to conduct follow-up observations on these pulsars. Student teams analyze these data and present their results on the final full day of the Student Institute.

In addition to research and pulsar-related activities, we expose the students to the various fields of expertise available at Green Bank. STEM disciplines are applied in all NRAO departments from the machine shop to scientific support. While on campus, students tour the GBT and lab areas including engineering, electronics, computer science, and central shops. They participate in seminars delivered by NRAO staff and summer undergraduate students.

West Virginia [WVU] students who participate in the PSC for a minimum of two years are eligible to receive WVU college credit in physics, and Virginia students can receive the same credit from the University of Virginia [UVa]. Students participating for two years may receive three credit hours, and students participating for three to four years may receive six hours of college credit.

### 2.1.2 *The Academic Year*

Teachers introduce pulsar astronomy and the PSC to whole classes of students through a series of inquiry-based activities developed during the Teacher Institute. The aim is to involve a larger group of students in elements of a research project and to interest students in joining local school teams.

Students who did not attend the Summer Institute but wish to join the PSC may do so after more training. They must demonstrate scientific rigor in order to provide scientists with reliable results since the results returned by the PSC are part of a real scientific research effort. To do so, they must pass two tests, which involve correctly distinguishing pulsars from RFI and noise in sample pulsar plots. Once the students join the PSC, they have full access to the PSC database and can start looking for new pulsars.

To analyze data, we have processed the raw data at a cluster located at WVU and placed the resulting diagnostic plots in our PSC database (http://psrsearch.wvu.edu). The data have been divided into *pointings*, small 150-second segments of the sky that correspond to the time it takes an astronomical source to drift through the field of view of the telescope at our observing frequency.  For each pointing, 35 diagnostic plots are created.  We have approximately 16,500 pointings, or 495,000 plots, for students to analyze, corresponding to thirty Terabytes of raw data and three hundred hours on the telescope.  The high-school students use the graphical database, shown in Figure 2, to quantitatively analyze the pulsar plots.

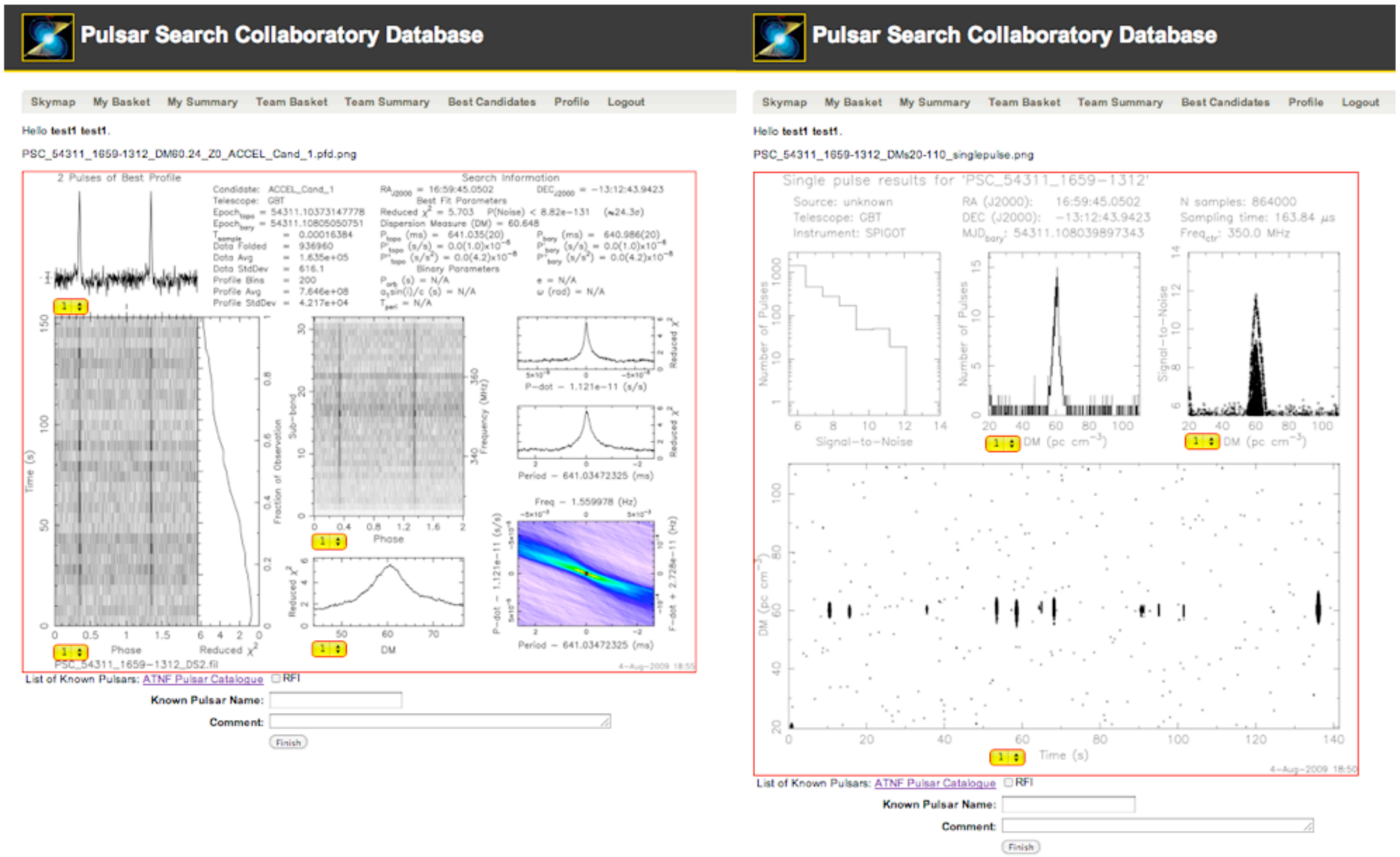

*Figure 2: A screenshot of the database used by the students to quantitatively rank the diagnostic pulsar plots.  These plots show the detection of a previously known pulsar, J1659-1305.*

Along with developing a database for the students to access the data, we developed a PSC website (http://www.pulsarsearchcollaboratory.com) for communication among the teachers, students, and astronomers and to provide students with supplemental information.  The site is used by students to discuss pulsar candidates among themselves and by teachers to monitor student progress within each individual school. The students also can communicate with students and teachers at other PSC schools. In addition, students can communicate with the astronomers through this site, and the astronomers can give useful tips and feedback to the students.  We also provide written material on how to analyze the pulsar plots (http://www.astro.virginia.edu/~rsl4v/PSC) and the details of pulsar searching (http://www.astro.virginia.edu/~rsl4v/PSC/presto_guide.pdf).

If a student finds a potential pulsar candidate, he or she posts his or her plot on the PSC website where the student receives prompt responses from the astronomers.  The students participate in follow-up observations of their candidates through Adobe

Connect Pro web conference software (http://www.adobe.com/products/acrobatconnectpro/demo). The software allows us to broadcast from the GBT control room. It also includes screen capture so that the students can watch the astronomers control the telescope, as well as a group chat window where the students can ask questions. If a student makes a discovery, we help her or him learn as much science as the student can from her or his object including: estimating its distance, measuring the spin-down rate to estimate its age and magnetic field, measuring its position and correlating it with catalogs from other telescopes, and establishing whether the pulsar has an orbiting companion.

#### 2.1.3 *Capstone Seminar*

Each year the PSC culminates in a three-day seminar at WVU. PSC teachers, students, staff, and school guidance counselors attend. During the seminar, PSC student members share their research via oral presentations, papers, or posters. Professional astronomers give talks and participate in discussions with the PSC students. WVU undergraduate and graduate students in physics, computer science, and engineering deliver short presentations on their various projects.

PSC students and guidance counselors broaden their awareness of STEM career pathways through tours of university science and engineering schools. Teachers, staff, evaluators, and Advisory Committee members participate in a capstone feed-back meeting at which the entire PSC community can reflect on successes and failures over the past year and share solutions and ways to improve the PSC for future years.

### 2.2 Evaluation of the Pulsar Search Collaboratory Program

To assess the progress of the PSC program, we implemented an evaluation plan comprised of several surveys. A variety of widely-used tests are available for assessing content knowledge, but we found that standard tests were not a good fit for our project since many concepts that are tested are not taught within the PSC. Instead, we developed a seven-item test that directly tests pulsar content knowledge, and we pre/post-test teachers and student leaders. In addition, we developed and validated a Likert scale instrument that gauges progress in self-efficacy and knowledge of the nature of scientific research. Teachers and students respond to statements that begin with “When I think of the Research Project, I feel” and are followed by various phrases to complete the sentence. Response choices on this Likert scale range from “Strongly Agree” to “Strongly Disagree.” Finally, we gather data on career interests of the students at various points in the academic year.

## 3. RESULTS AND DISCUSSION

### 3.1 What Students Are We Reaching?

Our project targets rural schools. Of the 25 schools in the PSC program, 11 schools are in rural school districts. Furthermore, we are reaching low-income students: One quarter of these students are from schools where over 50% participate in the Free/ Reduced School Lunch program.

We want to attract students irrespective of academic aptitude or prior interest in astronomy. Because the teachers select the students for the Student Institute, these students tend to be at the top of their class and have already expressed interest in STEM careers. Most (96%) plan to attend college. However, even though many of the students who attend the Student Institute are interested in STEM careers, only three of the nineteen students surveyed prior to the first Summer Institute expressed an interest in astrophysics as a future career choice. Students were interested in a range of fields for future careers, including accountant, aerospace engineer, architect, history teacher, hygienist, medical doctor, and radiologist.

*We note that the PSC project attracts students who could be first generation college-goers.* Of the students who attended the Student Institute, 26% reported that neither parent attended college. That percentage increased among students who joined the PSC during the academic year. Of the students who joined the PSC during the school year and attended the Capstone seminar, 40% reported that neither parent had gone to college.

### 3.2 Goal 1: Have we stimulated interest in STEM careers?

We asked students who attended the Student Institute and Capstone seminar to rate their interest in STEM careers after participating in the Summer Institute and Capstone Seminar. Overall, students show a significant increase in interest in STEM careers after participating in the PSC and especially in the Capstone Seminar, as shown in Figure 3. When we categorized the STEM fields by career (scientist, electrical engineer, software developer, mechanical engineer), we found that students were more interested in all of these STEM careers with the exception of mechanical engineering.

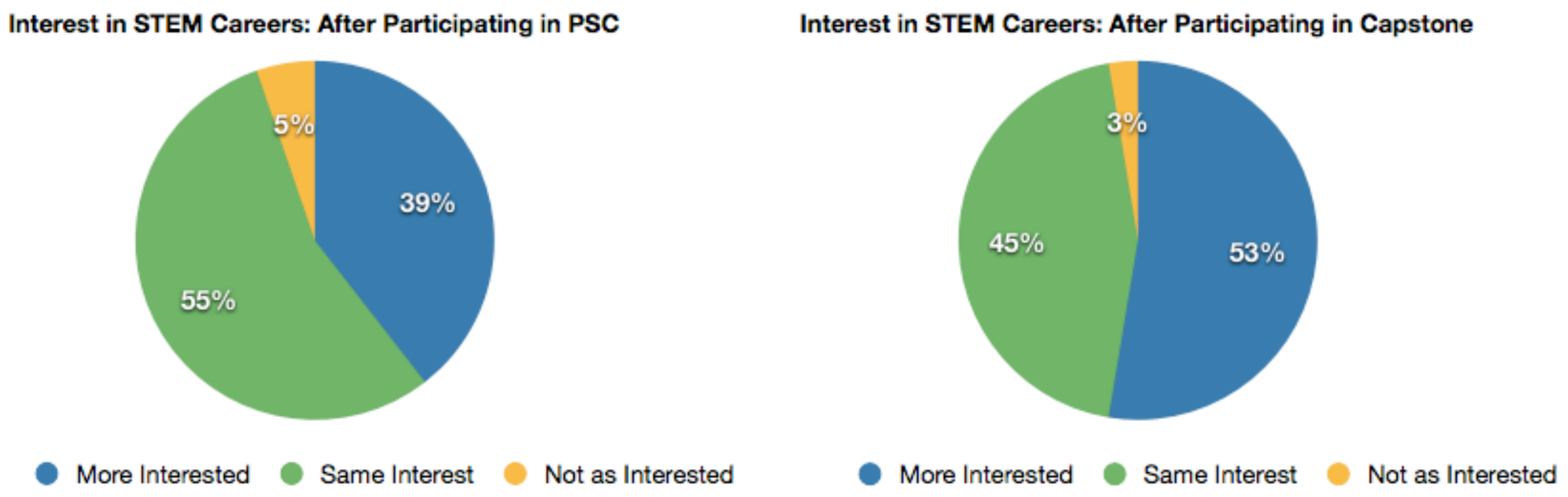

*Figure 3. The interest in STEM careers after participating in the PSC (left) and in Capstone Seminar (right).*

### 3.3 Goal 2: Have We Advanced Teachers and Students Understanding of the Nature of Science?

***Teachers***

To evaluate if teachers feel capable of conducting new research as a result of participating in the PSC, we asked them to take a Likert scale survey designed to gauge improvement in self-confidence and knowledge of the nature of scientific research. Whereas Cohort I teachers made pre/post gains at the $P < 0.1$ level, Cohort 2 teachers made significant pre/post gains at the $P<0.0009$ level on the total score of the scale for the survey and on 13 of the 27 items ($P<0.05$). This indicates that the changes we made to the Teacher Institute between the first and second years are effective.

| | **Table 1: Teachers: Research Self-Assessment** | | | | | | | | |
|---|---|---|---|---|---|---|---|---|---|
| | **Cohort 1 Teachers** | | | **Cohort 2 Teachers** | | | **Cohorts 1 and 2, Pooled** | | |
| | **N** | **Mean** | **Std. Dev.** | **N** | **Mean** | **Std. Dev.** | **N** | **Mean** | **Std. Dev.** |
| Pretest | 15 | 82.53 | 8.37 | 9 | 73.00 | 6.48 | 24 | 79.96 | 8.92 |
| Posttest | 15 | 87.67 | 10.55 | 9 | 84.67 | 9.10 | 24 | 86.54 | 9.93 |
| | One-sided paired t = 1.59<br>P < .0674 | | | One-sided paired t = 4.69<br>P < .0009 | | | One-sided paired t = 3.32<br>P < .0016 | | |
| | | | | | | | | | |

One teacher from Cohort 2 commented on a post-survey form: *I was totally "freaked out" initially – I didn't think I could do it... I wanted to crawl into a hole somewhere…I hated this survey…I've come a long way this week – I've pushed my own parameters with "group help."* Another teacher commented: *It was a great experience. Doing small research projects really helps so that the projects don't seem overwhelming.*

Teachers also showed a large improvement in their understanding of the scientific material. Quantitatively, the pre/post results on the seven-item pulsar assessment showed a significant improvement ($P=0.03$). Qualitatively, PSC staff noticed teachers' comprehensive grasp of pulsar science by the end of the Institute. The teachers presented the results of their research at the end of the Teacher Institute. The presentations were of high quality and, from the questions posed by teachers during the presentations, their understanding of the material was obvious.

***Students***

For students, chi-square analysis of each question in the Likert scale instrument shows significant ($P<0.05$) and improved differences in the distribution of responses for the following statements (pre/post), indicating that the PSC Institute succeeds in building confidence in students, rapport with the scientists involved in the project, and greater comfort with team-work. (See Figure 4.) The statements that show improvement are listed below.

When I think of the research, project, I feel:

• that I have the background I need for this research project;
• overwhelmed;
• afraid of making a fool out of myself;
• that I could do it better myself instead of with a team;
• that the scientists are too smart for me;
• afraid to ask the scientists questions;
• that if we don't replicate previous results it's no big deal.

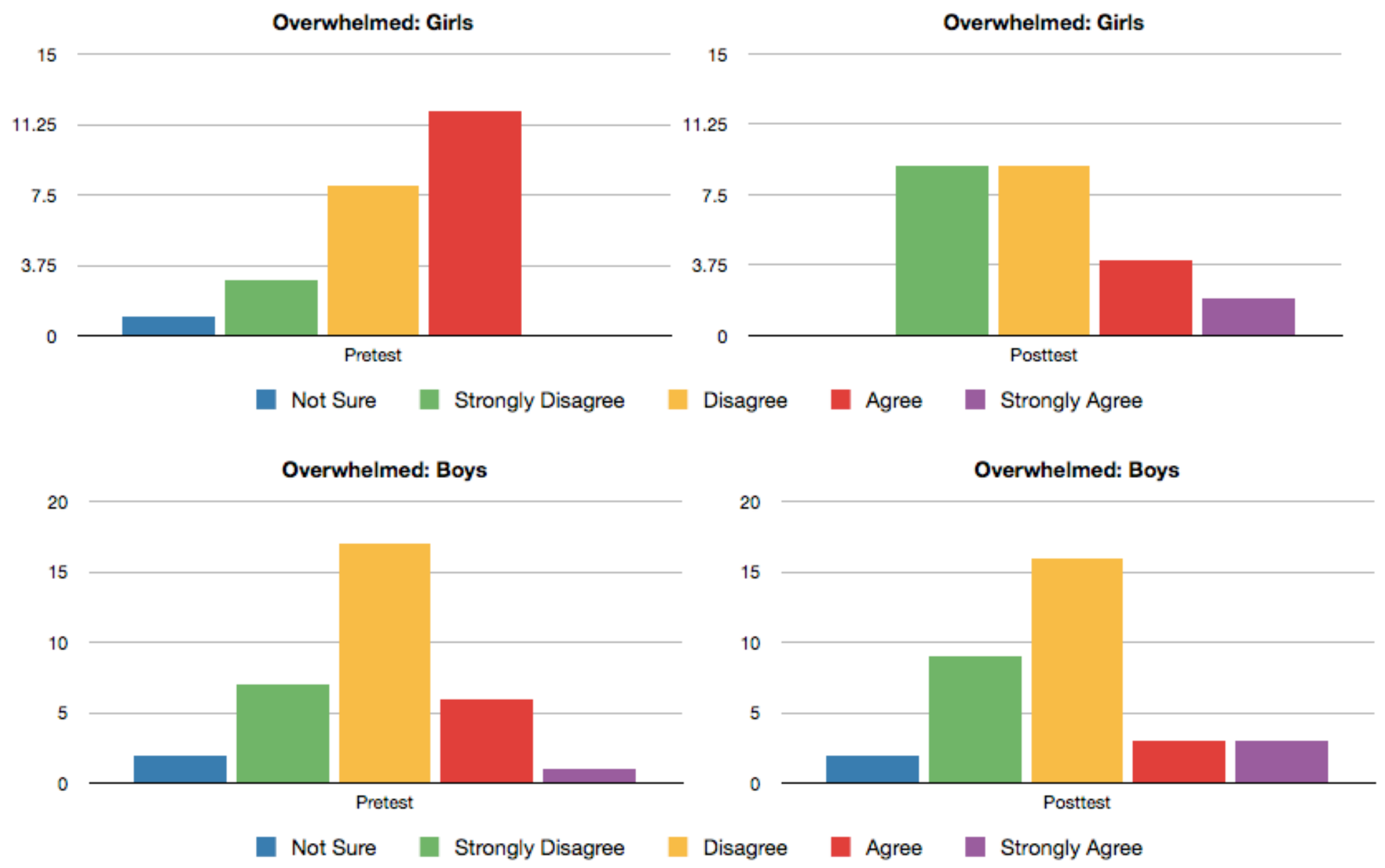

*Figure 4. The changes among girls (top) and boys (bottom) in their feelings of being overwhelmed before and after participation in the PSC.*

We see additional gains in girls, but not in boys, as they see themselves more as scientists after participating in the PSC program, as shown in Figure 5. This result is exciting and significant as self-efficacy is an important predictor of success in STEM fields (Clewell & Darke 2000). As a result of the PSC, girls:

• are more comfortable using scientific instruments;
• know how to answer a research question;
• see themselves as valuable members of a team;
• believe they will be doing valuable research.

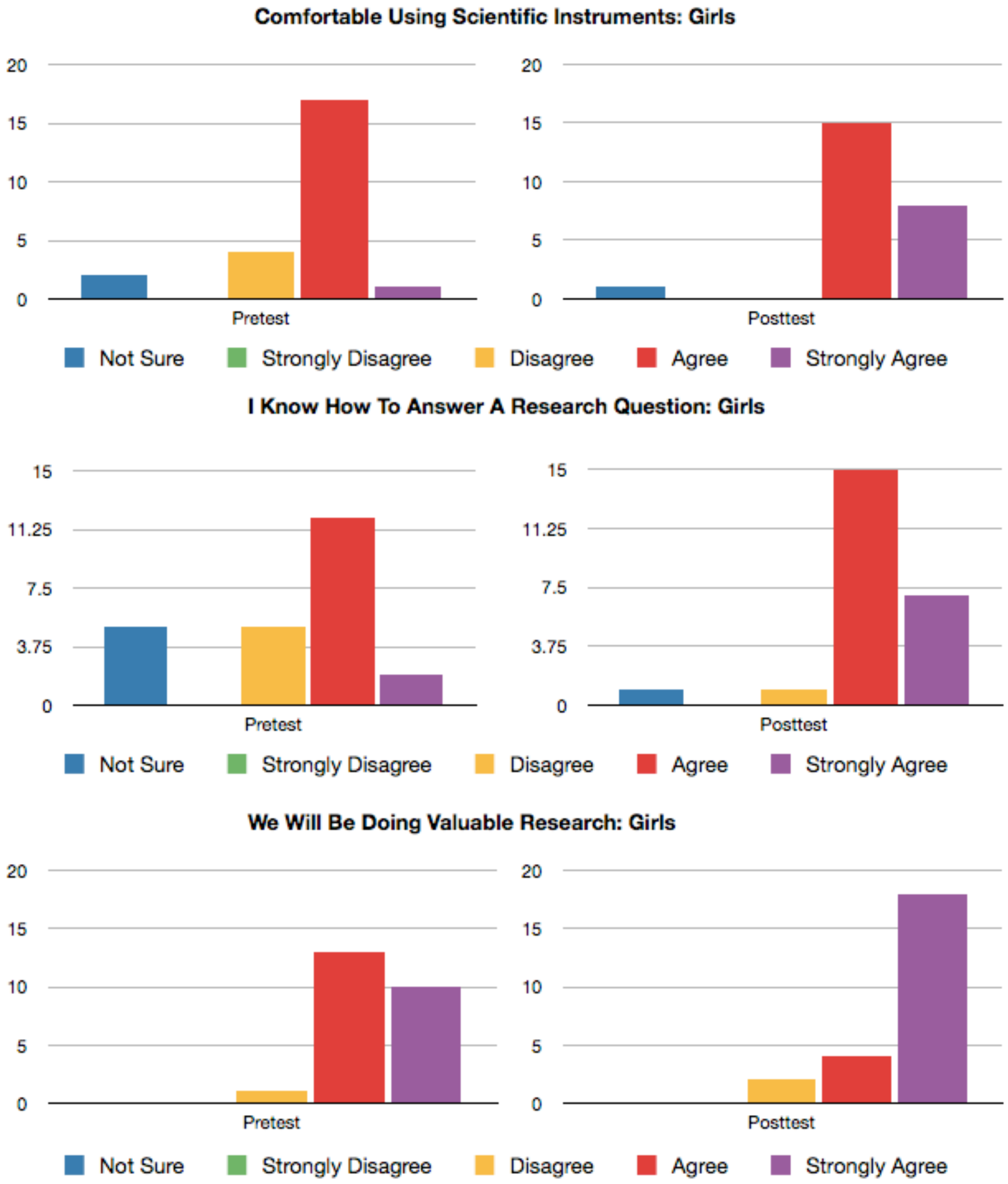

*Figure 5. The gains in self-efficacy in girls as a result of participation in the PSC.*

Like the teachers, students also present the results of their research at the end of the Student Institute. Students speak confidently and professionally about their results and demonstrate a clear understanding of pulsar research. The students are able to answer probing questions from the scientists and engineers in the audience and are able to devise future observations to try to answer remaining questions in their research.

### 3.4 Goal 3: Have We Promoted the Use of Information Technologies?

We assess the use of information technologies by two means: We monitor PSC website and database usage, and we survey the students at the Capstone Seminar.

To date, the 122 active students at 17 schools have examined 8,904 pointings on the PSC database. On average, these students spend 67 minutes on the database every week. Students post interesting plots to the PSC website for feedback from astronomers. Students have posted over 270 plots since the beginning of the PSC.

At the Capstone Seminar, students from Cohort 1 were surveyed regarding PSC website and database usage. Of the 38 students surveyed, 66% used the PSC e-mail to communicate with students from other schools, school teammates, and/or their teacher. Of these students, 53% used the PSC website chat feature for PSC related communication. In the thirty days prior to the Capstone Seminar, 32% of students used the database two-three times a week to analyze data, and 26% of students used the database once a week.

As part of the PSC, we use a variety of other online tools. For classes and groups that want or need additional instruction, we use Skype (http://www.skype.com) to communicate with the students. Based on feedback from teachers, Skype has a positive effect on the students: They are able to see and relate to the astronomers and feel that they are receiving the personalized attention that they need to be successful. To reach many students at once, we use the Adobe Connect software. For example, the students can participate in the follow-up observations of their candidates on the GBT using the Adobe Connect software, which allows video broadcast of the astronomer, two-way chat, and screen capture. We use this same software for on-line classes. We host miniature data-analysis marathons; we kicked off one marathon in January 2010 by broadcasting from the American Astronomical Society meeting in Washington, DC. In addition, the students regularly use the PSC database for scoring diagnostic pulsar plots. (Interestingly, one of the most well-received presentations at the Capstone Seminar was given by the database developer, a master's student at WVU; the students were fascinated about computer programming and the process of writing lines of software code that result in an on-line application that they use on a regular basis.) We also encourage students to research interesting plots online. For example, they can determine whether the distance to the candidate in their plot is a reasonable distance using the interactive tools at http://rsd-www.nrl.navy.mil/7213/lazio/ne_model.

### 3.5 Other Findings

While the program addresses content standards in physics and science, only a limited amount of time can be spent in the classroom on pulsar data analysis. Therefore, for most schools, the PSC has more in common with an informal science project and free-choice learning: Students choose to belong to the team, and most of the team meetings are held after school or during free time at school.

The chart in Figure 6, based on the responses of 38 students from Cohort 1 at the Capstone Seminar, shows where  and when most of the students conduct their data analysis.

| Number of Students | Description |
|---|---|
| 8 | After school at school |
| 5 | During school free time, after school at school, and after school at home |
| 4 | During school free time |
| 4 | After school at home |
| 3 | During school class time, school free time, at school, at home and another location |
| 3 | During school class time, school free time, and after school at home |
| 2 | During school class time, school free time, at school, and at home |
| 2 | During school class time and after school at home |
| 2 | After school at school and at home |
| 1 | During school class time and school free time |
| 1 | During school class time |
| 1 | During school free time, after school at school, home, and another location |
| 1 | During school free time and after school at home |
| 1 | After school at school and another location |
| TOTAL: 38 | |

*Figure 6.  Where the students conducted the majority of their data analysis for the PSC.*

Student retention in the second year of the PSC is surprisingly good. Of the 153 students who joined the PSC in year one, 58 were active members who qualified to attend the Capstone Seminar. Of that group, 6 students graduated, and 43 students (nearly 83%) from twelve schools are still active.

### 3.6  Scientific Results

Students have stayed engaged in the PSC, analyzing 8,904 pointings from our original data.  All of the pointings will be ranked by more than one team to build redundancy so that weak pulsars are not overlooked.  Of these 8,904 pointings ranked by students, 2,041 pointings are unique, and we have approximately 16,500 pointings in the original data.

To date, the PSC students have made two astronomical discoveries.  The first is a bright radio burst of astrophysical origin, most likely from a sporadic neutron star, and a pulsar. (See Afterward).  This bright radio burst, about 1.2 Janskys, was detected in the original data acquired during the summer of 2007 but has not been observed in the three twenty-thirty minute follow-up observations conducted in 2009.  The signal appears to originate about six hundred parsecs from the Earth.

A second student found evidence of an astronomical signal in a single pulse plot in October 2009.  Follow-up observations a month later led to the discovery of a 4.9-s pulsar at a distance of 1.4 kpc from the Earth.

The data acquired for the Pulsar Search Collaboratory project covers approximately 2,900 square degrees in the sky. In addition to new discoveries, students also are finding known pulsars. Students who find a potential pulsar in the data first check the ATNF catalog (http://www.atnf.csiro.au/research/pulsar/psrcat) to determine if a pulsar has been discovered in the same location in the sky with the same period and distance from the Earth. The PSC database keeps a list of the known pulsars that each student finds in the data. So far, the students have detected 23 known pulsars in the data that has been analyzed. If a student finds a known pulsar, they can estimate its distance, measure the spin-down rate to estimate its age and magnetic field, measure its position and correlate it with catalogs from other telescopes, check any misidentifications in previously catalogued parameters, and establish whether the pulsar has an orbiting companion.

## 4. CONCLUSIONS

Radio astronomy is an effective vehicle for interesting students in STEM careers, developing deeper understanding of the nature of science in teachers and students, and promoting the use of information technology by students and teachers. We also are having an enhanced effect on girls, and they are self-identifying themselves as scientists. Students constantly e-mail PSC astronomers with questions ranging from the details of analyzing pulsar plots to wanting to know what it is like to be an astronomer to how to build a cloud chamber to detect cosmic rays. Girls communicate with astronomers more frequently than boys. All questions get personal e-mail responses, and we encourage the students to keep asking questions. Perhaps, most significantly, three of the PSC leaders (the Education Director at NRAO, a faculty astronomer at WVU, and the Project Director/astronomer at NRAO) are female, offering girls not one, but *three*, role models. Each of these PSC leaders interacts with the students in one-on-one communication, and each plays a different role in the PSC, demonstrating by example that women can be successful in many areas of science and education.

When we asked students at the Capstone Seminar if they would recommend participation in the PSC to other students, 97% said they would. And all the teachers would recommend the PSC to their colleagues. These numbers are encouraging, as we have big plans for the PSC in upcoming years. The PSC has funding through July 2011, but we hope to sustain the project beyond that deadline. We will sustain the program beyond high school for students who elect to attend WVU. Astronomers at WVU are making plans to mentor the PSC students who attend WVU for college. A PSC@WVU club is in the works, designed to help students to both delve deeper into pulsar astronomy and to explore other areas of astronomical research. The club will include an observing center where students can conduct follow-up observations of new pulsars. Older students and graduate students also will take active roles in mentoring the less-experienced PSC students. We hope to have a positive impact on students not just in the near future but to support them in their journey along STEM career pathways for many years to come.

And finally, the long-term goal is to integrate PSC project tools and data into the National Virtual Observatory [NVO], making it possible for a broad cross-section of students and teachers across the nation, especially under-represented groups, to learn about the current practice of astronomy, actively participate in scientific research, and work with established astronomers.

## 5. AFTERWARD

A second discovery was recently made by Shay Bloxton, a West Virginia high-school sophomore. She first spotted evidence of a new pulsar on 15 October 2009. Bloxton, along with NRAO astronomers, observed the object again one month later. The new observation confirmed that the object is a pulsar, with a period of 4.9 seconds and a distance of 1.4 kpc from the Earth. Although Shay has not been invited to the White House, she is undeterred: “Participating in the PSC has definitely encouraged me to pursue my dream of being an astrophysicist," she says, adding that she hopes to attend West Virginia University to study astrophysics.

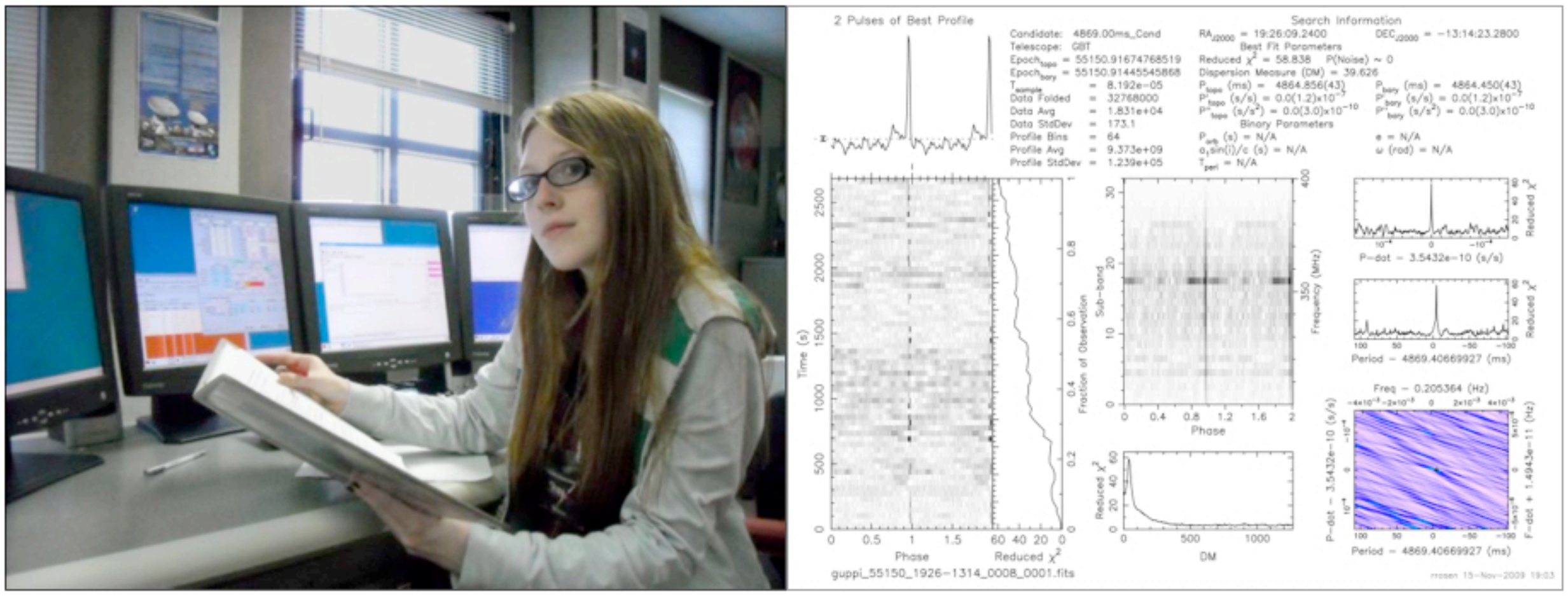

*Figure 7. Left: Shay Bloxton of Nicholas County High School, West Virginia. Right: The periodicity search output of the pulsar.*

While we know one astronomical discovery does not make a successful program, it does increase the excitement and motivation of our PSC students and the staff!

**Acknowledgements**

The Pulsar Search Collaboratory is funded by the National Science Foundation (award # 0737641). Partial support was provided by West Virginia EPSCoR. We would like to thank Don McLaughlin, Frances Van Scoy, and the West Virginia Virtual Environments Laboratory (http://wvvel.csee.wvu.edu) at WVU for their help processing the data. We would also like to thank Bill Saxton at NRAO for designing the PSC logos and Taylor Johnson at NRAO for designing the home page for the PSC database.